\documentclass[conference]{IEEEtran}
\IEEEoverridecommandlockouts

\usepackage{amsmath,amssymb,amsfonts}
\usepackage{graphicx}
\usepackage{textcomp}
\usepackage{xcolor}
\usepackage{xspace}
\usepackage{fancyhdr}
\usepackage{xfrac}
\usepackage{multirow}
\usepackage{adjustbox}
\usepackage[noend]{algpseudocode}
\usepackage{algorithm,algorithmicx}
\usepackage{hyperref}
\usepackage{color, colortbl}
\usepackage{comment}    

\excludecomment{arxiv}
\excludecomment{hide}

\definecolor{Gray}{gray}{0.9}

\newcommand{\Aydin}   [1]{{{\color{violet}(Ayd\i n) #1}}}

\newcommand{\Giulia}  [1]{{{\color{blue}(Giulia) #1}}}

\newcommand{\Oguz}[1] {{{\color{orange}(Oguz) #1}}}

\algdef{SE}[SUBALG]{Indent}{EndIndent}{}{\algorithmicend\ }%
\algtext*{Indent}
\algtext*{EndIndent}

\def\BibTeX{{\rm B\kern-.05em{\sc i\kern-.025em b}\kern-.08em
    T\kern-.1667em\lower.7ex\hbox{E}\kern-.125emX}}
    
\newcommand{\mA}{\mathbf{A}} 
\newcommand{\mC}{\mathbf{C}} 
 
\newcommand{\mR}{\mathbf{R}}
\newcommand{\mN}{\mathbf{N}} 
\newcommand{\mM}{\mathbf{M}}
\newcommand{\mS}{\mathbf{S}}
 
\newcommand{\dibella}{diBELLA\xspace} 
\newcommand{\mI}{\mathbf{I}} 
\newcommand{\vA}{\mathbf{v}} 
\newcommand{\transpose} {^{\mbox{\scriptsize \sf T}}}

\newcommand{\squared} {^{\mbox{\scriptsize \sf 2}}}
\newcommand{\kmer}[1][]{\textit{k}-mer#1}
    
\begin{document}

\title{Parallel String Graph Construction and Transitive Reduction for \emph{De Novo} Genome Assembly \\
}
\author{
    \IEEEauthorblockN{Giulia Guidi\IEEEauthorrefmark{1}\IEEEauthorrefmark{2}, 
    Oguz Selvitopi\IEEEauthorrefmark{2},
    Marquita Ellis\IEEEauthorrefmark{1}\IEEEauthorrefmark{2},
    Leonid Oliker\IEEEauthorrefmark{2},
    Katherine Yelick\IEEEauthorrefmark{1}\IEEEauthorrefmark{2},
    Ayd{\i}n Bulu\c{c}\IEEEauthorrefmark{1}\IEEEauthorrefmark{2} 
    }
    \IEEEauthorblockA{\IEEEauthorrefmark{1}Department of Electrical Engineering and Computer Sciences, University of California, Berkeley}
    \IEEEauthorblockA{\IEEEauthorrefmark{2}Computational Research Division, Lawrence Berkeley National Laboratory}
}

\maketitle

\begin{abstract}
One of the most computationally intensive tasks in computational biology is \emph{de novo} genome assembly, the decoding of the sequence of an unknown genome from redundant and erroneous short sequences.
A common assembly paradigm identifies overlapping sequences, simplifies their layout, and creates consensus.
Despite many algorithms developed in the literature, the efficient assembly of large genomes is still an open problem.

In this work, we introduce new distributed-memory parallel algorithms for overlap detection and layout simplification steps of \emph{de novo} genome assembly, and implement them in the diBELLA 2D pipeline. Our distributed memory algorithms for both overlap detection and layout simplification are based on linear-algebra operations over semirings using 2D distributed sparse matrices. Our layout step consists of performing a transitive reduction from the overlap graph to a string graph. We provide a detailed communication analysis of the main stages of our new algorithms.

diBELLA 2D achieves near linear scaling with over $80\%$ parallel efficiency for the human genome, reducing the runtime for overlap detection by $1.2\mbox{--}1.3\pmb\times$ for the human genome and $1.5\mbox{--}1.9\pmb\times$ for C.elegans compared to the state-of-the-art.
Our transitive reduction algorithm outperforms an existing distributed-memory implementation by $10.5\mbox{--}13.3\pmb\times$ for the human genome and $18\mbox{--}29\pmb\times$ for the C. elegans. 
Our work paves the way for efficient \emph{de novo} assembly of large genomes using long reads in distributed memory.


\end{abstract}

\begin{arxiv}
\end{arxiv}

\vspace*{1em}
\section{Introduction}\label{sec:introduction}

One of the greatest computational challenges for the analysis of high--throughput sequencing DNA fragments (namely \emph{reads}) is \emph{de novo} genome assembly~\cite{zhang2011practical}. 
It consists of aligning and merging redundant and incorrect DNA reads to reconstruct the original genome without any previous knowledge.

Long--read sequencing technologies~\cite{eid2009real, goodwin2015oxford} deliver sequences with an average length of more than 10,000 base pairs (bp).
The longer the sequences are read, the better. 
By using longer sequences, we can assemble through complex genomic repetitions to obtain more precise assemblies that were not possible with short--read technologies~\cite{phillippy2008genome,nagarajan2009parametric}.
Longer sequences come at the cost of higher error rates, which lead to higher algorithmic complexity and higher computational costs.

The Overlap--Layout--Consensus (OLC) paradigm is the most common assembly strategy for long--read data~\cite{berlin2015assembling}.
The first step (O) is to identify overlaps between reads to build an \emph{overlap} graph.
Due to the redundant sequencing and the inherent genome repetitiveness, the second step (L) simplifies the overlap graph and converts it into a \emph{string} graph.
A string graph is created from an overlap graph without contained edges and without transitive edges, where the edges represent the overlap \emph{suffix} and not the overlap itself.
Nevertheless, a string graph can be created from different source graphs depending on the application.
A string graph has the desirable property of collapsing genomic repeats into a single unit~\cite{simpson2012efficient}.
This conversion makes it easier to cluster sections of the graph into \emph{contigs}.
A contig is a set of overlapping sequences that together form a consensus region of DNA.
Then the consensus step (C) selects the most probable nucleotide sequence for each contig to correct errors in the data.
The OLC paradigm benefits from longer reads, since significantly fewer reads are required to cover the genome, limiting the size of the overlap graph.

Our earlier work~\cite{guidi2020bella, ellis2019dibella} focused on the implementation of parallel strategies for shared and distributed memory for the \emph{overlap} step.
In this respect, BELLA~\cite{guidi2020bella} is designed for shared memory and is the first work formulating overlap detection for \emph{de novo} genome assembly using sparse matrices. 
The distributed memory work~\cite{ellis2019dibella}, which we call diBELLA 1D, performs overlap detection using distributed hash tables.


In this work, we propose a sparse linear algebra centric approach called diBELLA 2D for distributed memory parallelization of overlap and layout phases.
By using 2D distributed sparse matrices for both phases, we reduce the need for different data structures in different steps of genome assembly. 
For the overlap step we formulate the overlap detection as a distributed Sparse General Matrix Multiply (SpGEMM).
For the layout step, we present a novel distributed memory algorithm for the \emph{transitive reduction} of the overlap graph.
This simplifies the overlap graph and makes it easier to resolve inconsistencies and create contigs.

A linear--time algorithm for the transitive reduction of an overlap graph~\cite{myers2005fragment} has been proposed earlier. However, that algorithm is inherently sequential.
By contrast, our transitive reduction algorithm is highly parallel.
Both the overlap and string graphs are represented as sparse matrices, and the entire transitive reduction algorithm is expressed as operations on sparse matrices.
In this direction, our contributions include the design of custom semirings, which are integral to the correctness of the algorithm.




Our results show the scalability of our pipeline for overlap detection plus transitive reduction and show that it achieves near linear scaling with over $80\%$ parallel efficiency for the human genome.
Our transitive reduction algorithm outperforms a competing distributed memory algorithm with a speedup of up to 13.3$\pmb\times$ for the human genome.
Our implementation is publicly available at \url{https://github.com/giuliaguidi/diBELLA.2D}. 




\begin{hide}
\Aydin{No need to waste space for the obvious}
\Giulia{Change url.}
The rest of this paper is organized as follows. Section~\ref{sec:background} describes the background necessary to understand the problem we are addressing, and Section~\ref{sec:related} presents the related work.
In Section~\ref{sec:methods} we describe our solution and its parallel implementation.
Finally, we evaluate our approach in Section~\ref{sec:results} and summarize our contributions in Section~\ref{sec:conclusions}.
\end{hide}

\section{Background}\label{sec:background}


A genome consists of one or more DNA molecules that are organized in three-dimensional space as \emph{chromosomes}.
The DNA consists of two sequences of nucleotides, called \emph{strands}, which wind around each other and form a double helix.
Each strand is a string over the alphabet $\Sigma = \{\rm A, C, G, T\}$ and has a direction.
The two strands of a DNA molecule have opposite directions.
On opposite strands, A always pairs with T and C with G.
One strand defines the \emph{reverse complement} of the other.
If $v = \rm ATTCG$, its reserve complement is $v' = \rm CGAAT$.
The \emph{canonical form} of a DNA sequence $v$ is the lexicographically smaller of $v$ and its reverse-complement $v'$. 
In our example, $v = \rm ATTCG$ is the canonical form.


OLC is the most widely used assembly paradigm for long read data~\cite{berlin2015assembling,li2016minimap}.
Its first step consists of identifying overlaps between input sequences.
The idea behind the search for overlaps is that two sequences that overlap may originate from adjacent positions on the genome.
However, the assembly process is more complex than it might seem at first glance.
This is because we cannot be sure that two overlapping sequences actually originate from adjacent positions on the genome due to the repetitiveness of the genome.

For the sequences $v_1$ and $v_2$ and their reverse-complements, $v_1'$ and $v_2'$ we can say that $v_1$ and $v_2$ have an overlap of length $L$ in base pair (bp) if and only if at least one of these is true:

\begin{itemize}
\item the last $L$ bp of $v_1$ match the first $L$ bp of $v_2$;
\item the last $L$ bp of $v_1$ match the first $L$ bp of $v_2'$;
\item the last $L$ bp of $v_1'$ match the first $L$ bp of $v_2$;
\item the last $L$ bp of $v_1'$ match the first $L$ bp of $v_2'$.
\end{itemize}

Given the erroneous sequencing process, in this context an overlap indicates that the two sequences have \emph{mostly} identical base content, so that their pairwise alignment score reaches a quality threshold defined by the overlap detection algorithm.

It is necessary to define four types of overlap, since reads can overlap in a reverse--complement manner and algorithms typically store only the canonical form of k-mers.
Additionally, we can define a \emph{contained overlap} as an overlap where the overlapping region of one read is the entire read.
An overlap can be called reverse--complement if and only if one of the sequences in the overlap is used in the original direction and the other one is used in the reverse--complement direction.
Therefore, an overlap that belongs to the last case is not a reverse--complement overlap, since it is equivalent to the last $L$ bp of $v_2$ matching the first $L$ bp of $v_1$.
Since the two forward cases are equivalent in theory, we only have three different cases.
However, it makes sense to keep the four categories in practice.
The correct use of orientation information is crucial during the second and third stages of the OLC algorithm and is essential for the correctness of the final assembly.

\begin{figure}
    \centering
    \includegraphics[width=\columnwidth]{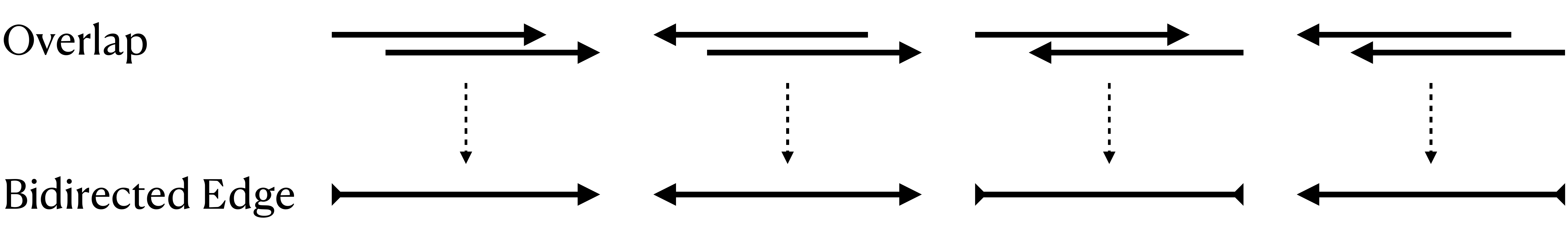}
    \caption{Overlap to bidirected edge type mapping.}
    \vspace{-1em}
    \label{fig:biedges}
\end{figure}

Commonly, an indexing data structure, such as a k-mer (i.e., a substring of fixed length $k$) index table or suffix array, is used to identify an initial set of overlap candidates~\cite{li2016minimap,koren2017canu,berlin2015assembling}.
Then, as a next step, pairwise alignment is sometimes performed to discard false positives.
After the overlaps have been calculated and consolidated from the reads, the next step is the \emph{layout} step, where the goal is to create a graph that encodes how we can assemble sequences to obtain contigs.

A string graph (or matrix) is a graph $G = (V, E)$, where $V$ is the set of sequences and $E$ is the set of overlap \emph{suffixes} between any two vertices.
There exists an edge if and only if the respective reads overlap and the weight of this edge is the length of the suffix.
For example, for the sequences $v_1 = \rm TACGA$ and $v_2 = \rm ACGACC$, their overlap suffix or \emph{overhang} is the portion of $v_2$ that exceeds the overlap between $v_1$ and $v_{2}$, i.e. $e_{12} = \rm CC$. 
Given $G = (V, E)$, where $V = \{v_1, v_2, v_3\}$ and $E = \{e_{12}, e_{13}, e_{23}\}$, we can walk (a) $v_1 \to v_2 \to v_3$ using $e_{12}$ and $e_{23}$, (b) or $v_1 \to v_3$ using only $e_{13}$.
If we take the weight of the edges into account (i.e., the overhang length), we can see that one of these two paths carries less information than the other and therefore we can mark it as \emph{transitive} and remove it from the string graph.

In \emph{de novo} genome assembly, we want to keep as many overlapping bases as possible for any pair of sequences, so we mark the edges (belonging to a valid path) with longer suffixes (higher weight) as \emph{transitive}.
Since the string graph maximizes the overlap length, it can disambiguate short repeats~\cite{simpson2012efficient}.
In our example, $e_{13}$ would be marked as \emph{transitive} and removed since the path $e_{12} \to e_{23}$ encloses more overlapping bases.

Since we do not know from which strand a certain sequence originates, we want to be able to traverse our graph in both forward and reverse direction.
That is, if we consider $G = (V, E)$ in our example above, we want to be able to walk both $v_1 \to v_2 \to v_3$ and $v_3' \to v_2' \to v_1'$.
Using the directed graph representation requires doubling the number of nodes, because for each read we need one vertex representing its \emph{entrance} and one representing its \emph{exit}.
Using an undirected graph can avoid this bloat, but it does not guarantee that a particular read will be used in a consistent manner at any point within a single assembly.
The use of a bidirected graph (i.e., a graph with a directional head at each end of each edge)~\cite{edmonds2003matching} solves both of these problems.
Figure~\ref{fig:biedges} shows the four types of bidirectional edges that result from the four overlap types described earlier.

\begin{figure}[t]
    \centering
    \includegraphics[width=\columnwidth]{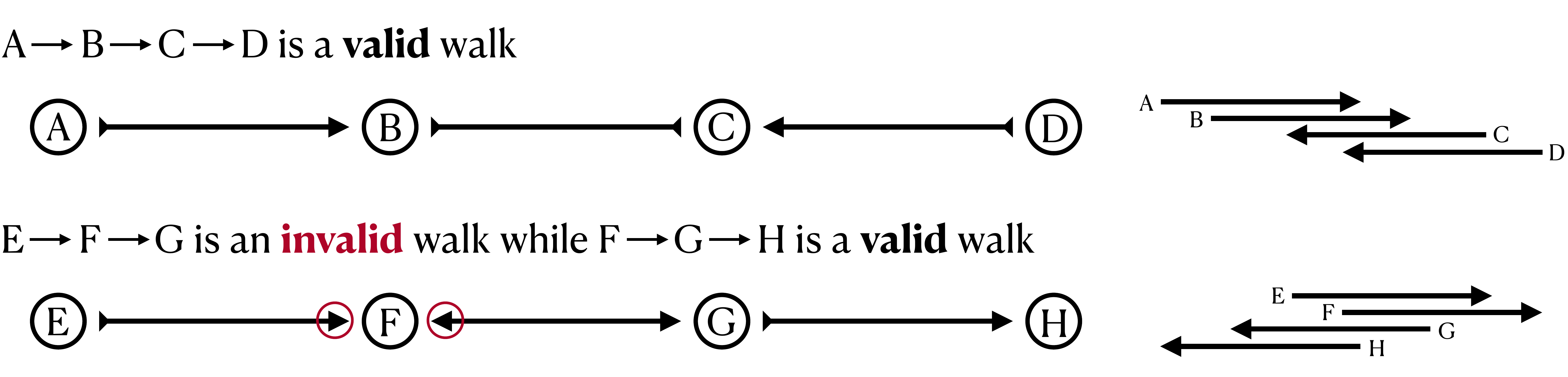}
    \caption{An example of a valid and invalid walk in a bidirected string graph.}
    \vspace{-1em}
    \label{fig:walk}
\end{figure}

If $G = (V, E)$ is a bidirected graph, then a valid path in $G$ is a continuous sequence of edges where each vertex is entered by a head inward and exited by a head outward (unless it is the end of the path) or vice versa.
Figure~\ref{fig:walk} shows two examples of a valid walk ($A \to B \to C \to D$ and $F \to G \to H$) and one example of an invalid walk ($E \to F \to G$).
A walk through the bidirected string graph encodes the way the sequences can be consistently assembled~\cite{myers2005fragment}. 

Given two paths $v_1 \to v_2 \to v_3$ and $v_1 \to v_3$ in a bidirected string graph, the edge $v_1 \to v_3$ can only be considered to be transitive if the following conditions are satisfied: (a) $v_1 \to v_2 \to v_3$ constitutes a valid walk, (b) the two heads next to $v_1$ have the same orientation, and (c) the two heads next to $v_3$ also have the same orientation.

By definition, a string graph can be constructed from various sources, such as an overlap graph (as we present in this paper), k-mers~\cite{jackson2008parallel} or FM index~\cite{bonizzoni2017fsg}, and Burrows--Wheeler transform (BWT)~\cite{simpson2012efficient}.
These approaches are not invariant to the input properties and often only consider error-free sequences. 
In reality, long read data with its high error rates and long lengths often make string graph construction impractical for approaches other than those based on overlap graphs.

\section{Related Work}\label{sec:related}

In this section we review the literature on overlap detection and transitive reduction, and describe works related to ours.

Myers' transitive reduction algorithm consists of iterating over each node $v$ in the source graph and examining nodes up to two edges away from $v$ to identify all transitive edges that leave or enter $v$~\cite{myers2005fragment}.
These edges are then marked for removal, and they are removed after all nodes have been considered. 

Li~\cite{li2018minimap2,li2016minimap} uses a seed--based approach to find overlaps and then uses the Myers algorithm to transitively reduce the string graph.
In particular, Li uses \emph{minimizers} (i.e., reduced k-mer representations) for overlap detection and does not perform explicit pairwise alignment on sequences.

Simpson and Durbin~\cite{simpson2012efficient} use the Ferragina--Manzini index (FM--index)~\cite{ferragina2005indexing} derived from the Burrows--Wheeler transform~\cite{burrows1994block} for overlap detection and transitive reduction. 
Their algorithm for transitive reduction is similar to that of Myers'.
Bonizzoni et al.~\cite{bonizzoni2017fsg} propose a similar approach using only the FM--index of the input sequence to create a string graph and perform a transitive reduction with a different but equivalent formulation than Myers'~\cite{bonizzoni2017external}.

BELLA~\cite{guidi2020bella} is the first work proposing to use sparse matrices for overlap detection.
A sparse matrix $\mA$ is used to indicate the occurrence of k-mers in sequences, and the multiplication of this matrix by its transpose, i.e. $\mA\mA\transpose$, is used to detect the overlaps.
Given the similarity of our distributed memory design to BELLA, we give more details about it in Section~\ref{sec:methods}. 


Our first distributed memory design for overlap detection, diBELLA 1D~\cite{ellis2019dibella}, uses a k-mer based approach and creates and traverses a distributed hash table to find overlapping sequences.
\begin{arxiv}
The keys of this hash table are the k-mers and the values are lists of sequence identifiers in which the respective k-mer appears.
\end{arxiv}
This design resembles a 1D SpGEMM using an outer product algorithm without explicit construction of matrices.
It does not perform transitive reduction.

PASTIS~\cite{selvitopi2020pastis} is inspired by BELLA and computes protein homology search as distributed SpGEMM.
Genome assembly and protein homology search are different problems, but both require a computationally expensive all-to-all comparison.

Besta et al.~\cite{besta2020communication} present another approach similar to BELLA using distributed SpGEMM to calculate the Jaccard similarity between read sets of different genomes.
The main difference is that their software is optimized for the case where the output $\lvert\textit{genomes}\rvert$-by-$\lvert\textit{genomes}\rvert$ matrix is dense because it stores the Jaccard similarity between any genome pairs.

Jackson and Aluru~\cite{jackson2008parallel} present a parallel algorithm for constructing a bidirected string graph from a de Bruijn graph~\cite{medvedev2007computability} in which the vertices represent k-mers and edges correspond to individual nucleotides whose two vertices have in common.
A de Bruijn graph is not suitable for long read data because of the high error rates.

SORA~\cite{paul2018sora} computes transitive reduction of a string graph based on an overlap graph in distributed memory using Apache Spark~\cite{zaharia2016apache} and the GraphX library~\cite{gonzalez2014graphx}, which allows parallel computation on distributed graphs in Spark.
To the best of our knowledge, SORA is the only other distributed algorithm that computes transitive reduction on overlap graphs, although it was designed for cloud environments.
\section{Proposed Algorithm}\label{sec:methods}

In this section we describe the current implementation of our pipeline, focusing primarily on the novel transitive reduction algorithm.
To keep the paper self--contained, we first briefly describe the first half of the pipeline, which combines k-mer counting and overlap detection of our prior work~\cite{georganas2015hipmer, selvitopi2020pastis}.


\vspace{-.25em}
\subsection{Overview}

\begin{algorithm}[t]
    \caption{The matrix computation in diBELLA 2D.}
    \label{alg2}
    \begin{algorithmic}[1] 
        \Procedure{diBELLA 2D}{}
             \State $\textit{reads} \gets\Call{FastaReader}$ 
             \State $\textit\kmer{s} \gets \Call{KmerCounter}$
             \State $\mA \hspace{0.43em}\gets \Call{GenerateA}{\textit{reads},\hspace{0.0825em}\textit\kmer{s}}$\Comment{Data matrix}
             \State $\mA\transpose \gets \Call{Transpose}{\mA}$
             \State $\mC \gets \mA\mA\transpose$\Comment{Candidate overlap matrix}
             \State $\mC \gets \Call{Apply}{\mC, \rm Alignment()}$\Comment{Run alignment}
             \State $\mR \gets \Call{Prune}{\mC, \rm AlignmentScoreLessThan(\textit{t})}$
             \State $\mS \gets \Call{TransitiveReduction}{\mR}$\Comment{Algorithm~\ref{alg1}}
            \State \textbf{return} $\mS$
        \EndProcedure
    \end{algorithmic}
\end{algorithm}

Our algorithm design~\cite{guidi2020bella} uses a k-mer--based approach and relies on parallel sparse matrix multiplication for overlap detection as the first step of the OLC paradigm. 
The outline of our pipeline can be seen in Algorithm~\ref{alg2}.
We form a $\lvert\textit{sequences}\rvert$-by-$\lvert\textit\kmer{s}\rvert$ matrix $\mA$ to detect the occurrence of k-mers in sequences, and perform $\mA\mA\transpose$ to detect overlaps, resulting in a sparse $\lvert\textit{sequences}\rvert$-by-$\lvert\textit{sequences}\rvert$ matrix $\mC$.
Overlap detection is followed by a computationally intensive seed--and--extend pairwise alignment for all nonzeros in $\mC$ using SeqAn~\cite{doring2008seqan}, a sequence analysis library. 
If the alignment score of a read pair does not exceed a threshold, then the overlap is discarded and the entry is removed from the matrix.
We refer to this resulting output $\lvert\textit{sequences}\rvert$-by-$\lvert\textit{sequences}\rvert$ matrix as \emph{overlap} matrix and denote it with $\mR$.
A transitive reduction algorithm is then run on $\mR$ to remove redundant edges and simplify it into a string graph, $\mS$.


\vspace{-.25em}
\subsection{Data Partitioning}
\label{sec:data-part}
The input to our program is a set of nucleotide sequences in FASTA file format.
To ensure load balance, each processor reads an equal--sized independent chunk of this file via parallel MPI I/O.
Immediately thereafter, processors begin communicating sequences to create a 2D grid that is consistent with the way the matrices are partitioned among processors.
This approach is similar to the one chosen by PASTIS~\cite{selvitopi2020pastis}.

\vspace{-.25em}
\subsection{K-mer Selection and Counting}

A k-mer based approach calculates the frequency of each k-mer in the input because not all k-mers are useful.
K-mers that are usually discarded are (a) k-mers that occur only once in the input (\emph{singletons}) and (b) high frequency k-mers.
For more details on k-mer selection, see BELLA paper~\cite{guidi2020bella}.



\dibella 2D eliminates singletons using a Bloom filter~\cite{melsted2011efficient} during k-mer counting and high frequency k-mers that occur at least $d$ times, as in our first implementation.
The threshold $d$ is calculated using the approach introduced in BELLA~\cite{guidi2020bella}, which uses dataset-specific features.
In Section~\ref{sec:comm} we use $d$ to calculate the communication costs of our algorithm.

\begin{figure}
    \centering
    \includegraphics[width=0.85 \columnwidth]{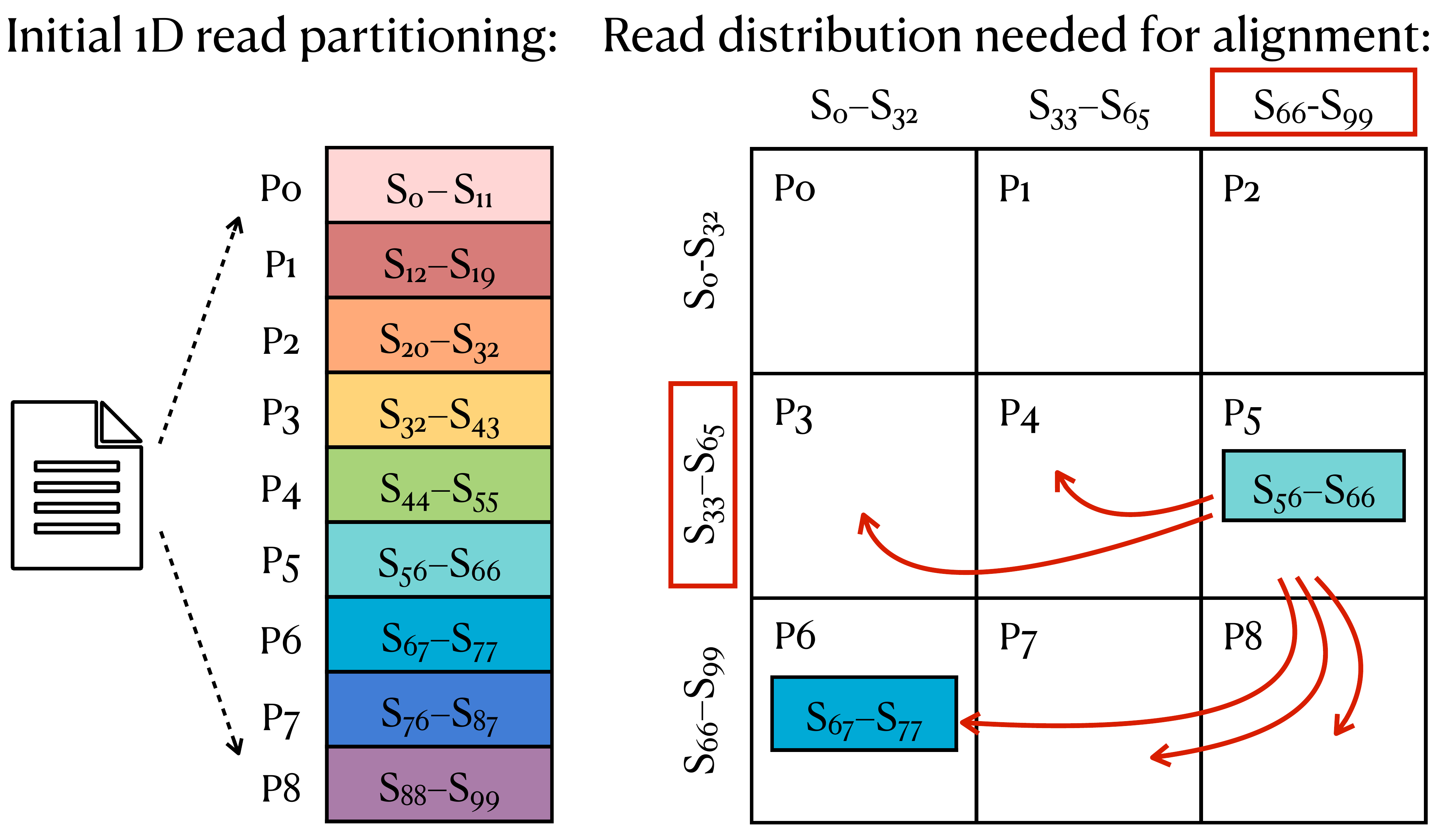}
    \caption{Read distribution when we read the input (left) and read distribution we need to perform pairwise alignment (right).}
    \vspace{-1em}
    \label{fig:data_partition}
\end{figure}

Our k-mer counter is similar to that of HipMer~\cite{georganas2015hipmer} and consists of two phases.
First we add k-mers to the Bloom filter and then we calculate the frequencies for the filtered k-mers.
The processors extract k-mers from their local sequences, hash them, and possibly communicate them with other processors as dictated by the Bloom filter hash function.
On the receiver, the incoming k-mers are added to the local Bloom filter; if they already exist, they are added to the local hash table partition.
The communication requires an all-to-all exchange and is implemented via MPI Alltoall and MPI Alltoallv.

\vspace{-.25em}
\subsection{Overlap Detection and Alignment}

The local k-mer hash table and the local sequences are used to create a distributed $\lvert\textit{sequences}\rvert$-by-$\lvert\textit\kmer{s}\rvert$ matrix $\mA$.
A nonzero $\mA_{ij}$ stores the position of the $j$th k-mer in the $i$th sequence.
$\mA$ is multiplied by $\mA\transpose$ to obtain the sparse candidate overlap matrix $\mC = \mA\mA\transpose$ of dimension $\lvert\textit{sequences}\rvert$-by-$\lvert\textit{sequences}\rvert$.
In $\mC$ each nonzero $\mC_{ij}$ stores the number of common k-mers and their positions in the sequence pair $i$ and $j$.
The number of stored positions is a user-defined parameter.
For this work we store two k-mer positions for each read pair.

To compute $\mC$ we use the distributed SpGEMM in the CombBLAS library~\cite{bulucc2011combinatorial} and we overload the addition and multiplication operators in SpGEMM with a custom semiring.
We overload the multiplication with an assignment by taking the positions of the respective k-mer in two sequences corresponding to $\mA_i$ and $\mA\transpose_j$.
We overload the addition operator by incrementing the counter of common k-mers between $\mA_i$ and $\mA\transpose_j$ and storing the positions of another common k-mer in $\mC_{ij}$ (i.e., concatenate the results of the multiplication operation) as long as it is smaller than the number of positions to be stored.

CombBLAS relies on 2D Sparse SUMMA algorithm for parallel SpGEMM~\cite{buluc2008challenges} and it uses a hybrid hash table and heap based algorithm for local multiplication.
For matrices, CombBLAS uses a 2D matrix decomposition, so both $\mA$ and $\mC$ are distributed over a process grid of $\sqrt{P}\times \sqrt{P}$.
Observe that in such a distribution a processor may need to align a pair of sequences which it does not have in its partition (Figure~\ref{fig:data_partition}).
Such sequences need to be communicated among respective processors.
In this respect, a processor has two possibilities: (a) wait until $\mC$ is computed to find out which sequences it would need, and then begin communicating those sequences, or (b) request the full range of sequences it might need once the FASTA input file is read from disk, as described in Section~\ref{sec:data-part}.
We choose the latter option because it allows for overlapping sequence exchanges with k-mer counting and matrix multiplication.
This is also the approach adopted by PASTIS~\cite{selvitopi2020pastis}.
Once the sequences are communicated, we perform pairwise alignment for all identified pairs (i.e., the nonzeros $\mC$) using a seed--and--extend algorithm that returns an alignment score and updated seed coordinates.
If the score does not exceed the specified threshold, the read pair is discarded and the nonzero is removed from $\mC$.

We can view these operations from a matrix point of view: (a) the pairwise alignment is an in-place element-wise operation on $\mC$ that sets the alignment flag to true for each nonzero $\mC_{ij}$ if the alignment score exceeds the given threshold, and to false otherwise, and (b) the removal of entries with false flags is another in-place operation on $\mC$ that prunes nonzeros whose flags are set to false.
The resulting matrix $\mR$ (line 8 in Algorithm~\ref{alg2}) at the end of these operations is the input for the transitive reduction algorithm.
Contained overlaps, as defined in Section~\ref{sec:background}, are discarded during transitive reduction regardless of their alignment scores. 
They may be reintroduced at later stages of the \emph{de novo} assembly process.


\vspace{-.25em}
\subsection{Transitive Reduction}


Our distributed memory transitive reduction algorithm takes the overlap matrix $\mR$ as input and computes a transitively reduced version of $\mR$, which we refer to as $\mS$ (line 9 in Algorithm~\ref{alg2}).
Recall that each nonzero $\mR_{ij}$ stores the number of common k-mers and their positions in the sequence pair $(s_i,s_j)$.
The transitive reduction algorithm needs two additional information for each such pair: the length of the overlap suffix and the overlap orientation.
Both can be derived in-place from the alignment coordinates stored in $\mR_{ij}$.
\begin{arxiv}
\Oguz{(you do not seem to be using $s^E$ anywhere, remove it?)}For a sequence $s$, let $s^{S}$ denote the start position of the alignment and $s^{E}$ denote the end position of it.
Given the set of overlapping sequence pairs $(s_i, s_j)$, we divide them into four categories and assign them an orientation as follows \Oguz{(the below categorization is also not utilized anywhere in this section)}:
\begin{itemize}
    \item If $s_i, s_j$ are \textbf{not} reverse-complement and $s_i^{S} > s_j^{S}$, we label it as a \emph{forward overlap};
    \item If $s_i, s_j$ are \textbf{not} reverse-complement and $s_i^{S} < s_j^{S}$, we label it as a \emph{reverse overlap};
    \item If $s_i, s_j$ are reverse-complement and $s_i^{S} > s_j^{S}$, we label it as an \emph{inner overlap};
    \item If $s_i, s_j$ are reverse-complement and $s_i^{S} < s_j^{S}$, we label it as an \emph{outer overlap}.
\end{itemize}
\end{arxiv}
The length of the suffix and the orientation are calculated and stored in $\mR$ during the pairwise alignment, so that $\mR$ is immediately ready for the transitive reduction phase.

Our transitive reduction algorithm is presented in Algorithm~\ref{alg1}.
The algorithm begins by discovering two-hop neighbors of each vertex in the overlap graph.
This first step is the most computationally intensive stage of transitive reduction and is achieved by squaring the overlap matrix: $\mN = \mR\squared$, where $\mN$ is the two-hop \emph{neighbor} matrix.
In \emph{de novo} assembly, whenever there are multiple alternative paths in the graph, we retain the one that gives us more genomic coverage in terms of nucleotides.
Therefore, whenever there are multiple alternatives, the path with the shorter suffix is chosen, since a shorter suffix indicates a longer overlap between two sequences.
This is achieved by using a custom MinPlus semiring during the squaring of $\mR$. 
Algorithm~\ref{alg-min} illustrates the MinPlus semiring we use, where we overload the addition operation with a minimum operation and the multiplication operation with a summation.
Because of the bidirectionality of our graph, we make sure that the orientation of the edges in play conforms to the transitivity rules listed in Section~\ref{sec:background}.
This is ensured by checking whether the edges follow the transitivity rules during multiplication (line 5 in Algorithm~\ref{alg-min}). 
If not, we mark the edge as \emph{not transitive}.
In particular, we check in MinPlus semiring whether the two heads next to the intermediate node (i.e. the middle node of a three-node path) have opposite directions.
\begin{arxiv}
For example, a forward overlap and an inner overlap are okay, while a forward overlap and an outer overlap mark the two-hop edge as non-transitive.
\end{arxiv}

\algdef{SE}[DOWHILE]{Do}{doWhile}{\algorithmicdo}[1]{\algorithmicwhile\ #1}%

\begin{algorithm}[t]
    \caption{Parallel transitive reduction on $\mR$.}
    \label{alg1}
    \begin{algorithmic}[1] 
        \Procedure{TransitiveReduction}{$\mR$}
            \Do
                \State $prev \gets \mR.\Call{nnz}{}$
                \State $\mN \gets \mR\squared$\Comment{Find edges two-hops away}
                \State $\hspace{0.0825em}\vA\hspace{0.0825em} \gets \mR.\Call{Reduce}{\mathsf{Row}, 0, \mathrm{max}}$ 
                \State $\hspace{0.0825em}\vA\hspace{0.0825em} \gets \vA.\Call{Apply}{x, \mathrm{ add}}$\Comment{\emph{x} is a scalar}
                \State $\mM \gets \mR.\Call{DimApply}{\mathsf{Row},\vA,\mathrm{ return2nd}}$
                \State $\hspace{.225em}\mI\hspace{.225em} \gets \mM \geq \mN$\Comment{Find transitive edges}
                \State $\hspace{.1em}\mR \gets \mR \circ \neg\mI$\Comment{Remove transitive edges}
                \State $nnz \gets \mR.\Call{nnz}{}$
            \doWhile{$nnz \neq prev$}
            \State $\mS \gets \mR$
            \State \textbf{return} $\mS$
        \EndProcedure
    \end{algorithmic}
\end{algorithm}
\algrenewcommand\algorithmicprocedure{\textbf{struct}}
\begin{algorithm}[t]
    \caption{Custom MinPlus semiring used in $\mN \gets \mR\squared$.}
    \label{alg-min}
    \begin{algorithmic}[1] 
        \Procedure{MinPlusSR}{}{}
            \State $\Call{Id(\hspace{.15em})}{}~\mathrm{\textbf{return}}~{\infty}$
            \State $\Call{Add}{a, b}~\mathrm{\textbf{return}}~\Call{Min}{a,b}$\Comment{Find the shortest path}
            \State $\Call{Multiply}{a, b}$
            \Indent
                \If {$\Call{IsDirOK(\hspace{.15em})}{}$}~\textbf{return} $a+b$
                \Else ~\textbf{return} $\Call{Id(\hspace{.15em})}{}$
                \EndIf
            \EndIndent
        \EndProcedure
    \end{algorithmic}
\end{algorithm}

\begin{arxiv}
Let us assume the following is our overlap matrix $\mR$, where zeros are actually zeros in the matrix (i.e. non-existent entries) and nonzeros represent the overlap suffix.
To keep the running example simple, also assume that edges only exist in the forward direction.

\[
\mR =
\begin{bmatrix}

0  & 0  & 0  & 0    \\
30 & 0 & 0 & 0      \\
80 & 30 & 0 & 0     \\
0 & 80 & 30 & 0 

\end{bmatrix},~
\mN = \mR\squared =
\begin{bmatrix}
0  & 0  & 0  & 0   \\
0 & 0 & 0 & 0      \\
60 & 0 & 0 & 0     \\
30 & 60 & 0 & 0 
\end{bmatrix}
\]

In $\mR$ we can go from the first to the second read, from the second to the third, and from the third to the fourth with overlap suffixes of length 30.
But we can also go from the first to the third with a suffix of length 80, and in a similar way from the second to the fourth.
Our objective is to preserve the path with shortest edges, therefore we want to remove the two edges of length 80 from $\mR$.

In $\mN$ we identify the shortest path from a particular read to its two hop neighbors.
The shortest way from the first to the third read has a length of 60, since it passes through two edges of length 30. 
The path from the first to the third read in a single hop would cost 80, and we want to minimize rather than maximize this distance.
\end{arxiv}

In lines 5--7 of Algorithm~\ref{alg1}, we create the \emph{maximal suffix} matrix $\mM$, where each nonzero within a row is replaced by the maximum value (i.e., the longest overlap) of that row.
In \emph{de novo} assembly, read overlaps are approximate matches because sequencing errors can cause endpoint positions to shift.
To make our algorithm robust to sequencing errors, we increase the value of the longest overlap per row (i.e. per read) by a scalar $x$. 
$\Call{Reduce}{}$ in line 5 returns a vector $\vA$ whose $i$th cell stores the maximum value of the $i$th row of $\mR$.
$\Call{Apply}{}$ in line 6 adds $x$ to each nonzero $\vA_i$.
For each nonzero $\vA_i \neq 0$, $\Call{DimApply}{}$ in line 7 replaces the corresponding row $i$ of $\mR$ with $\vA_i$ in its output. 
The last parameter in the functions $\Call{Reduce}{}$, $\Call{Apply}{}$ and $\Call{DimApply}{}$ is the binary operator applied to each scalar operation.


\begin{arxiv}
In our example we assume that $x = 10$, so we have:

\[
\vA =
\begin{bmatrix}

0      \\
30     \\
80     \\
80 

\end{bmatrix},~
\vA = \vA + x =
\begin{bmatrix}
0          \\
40     \\
90     \\
90 
\end{bmatrix},~
\mM =
\begin{bmatrix}

0  & 0  & 0  & 0    \\
40 & 0 & 0 & 0      \\
90 & 90 & 0 & 0     \\
0 & 90 & 90 & 0 

\end{bmatrix}
\]
\end{arxiv}

The next step is to identify the transitive edges in $\mN$ (line 8 in Algorithm~\ref{alg1}).
Our algorithm performs an element-wise operation between $\mM$ and $\mN$ to identify such edges.
If $\mM_{ij}$ is greater than or equal to $\mN_{ij}$, the corresponding nonzero $\mI_{ij}$ in the output matrix is set to true.
In $\mN$ we store the shortest path, so that all nonzeros with $\mM_{ij} \geq \mN_{ij}$ are transitive edges because $\mM_{ij}$ is an overlap suffix longer than $\mN_{ij}$.

\begin{arxiv}
In our running example the output matrix is $\mI$:

\[
\mI = \mM \geq \mN = 
\begin{bmatrix}

0  & 0  & 0  & 0    \\
0  & 0  & 0  & 0    \\
1  & 0  & 0  & 0    \\
0  & 1  & 0  & 0 
\end{bmatrix}
\]
\end{arxiv}

The described element-wise operation is only performed for entries that are nonzero in both $\mM$ and $\mN$.
Recall that when computing $\mN = \mR\squared$ we checked whether a path is a valid walk or not. 
In this element-wise operation we also make sure that the orientation of the edges in the intersection of $\mN$ and $\mM$ follows the last two transitivity rules.
In the element-wise operation, we check whether the two heads next to the \emph{departure} node (i.e., the start node of a three-node path) and the two heads next to the \emph{destination} node (i.e., the end node of a three-node path) have the same orientation.

The final operation of our transitive reduction algorithm is to prune the identified transitive edges from $\mR$ (line 9 in Algorithm~\ref{alg1}).
This is achieved by an element-wise multiplication of $\mR$ and the logical negation of $\mI$, $\neg\mI$.
Any nonzero in $\mI_{ij}$ becomes a zero in $\neg\mI_{ij}$, therefore the nonzeros of $\mR$ corresponding to the transitive edges (i.e. zeros) in $\neg\mI$ are pruned.
Again, only those entries that are nonzero in both $\mR$ and $\neg\mI$ are considered.
This is equivalent to a set difference operator ($\mathit{nonzeros}(\mR)\setminus\mathit{nonzeros}(\mI))$ in linear algebra.

\begin{arxiv}
In our example, $\mR$ becomes:

\[
\mS = \mR \circ \neg\mI = 
\begin{bmatrix}

0  & 0  & 0  & 0    \\
30  & 0  & 0  & 0    \\
0  & 30  & 0  & 0    \\
0  & 0  & 30  & 0 
\end{bmatrix}
\]

In this example, all transitive edges are removed after only one iteration of the algorithm and so we get $\mS$.
\end{arxiv}

In practice, we need several rounds to successfully remove all transitive edges, since we need to consider neighbors that are three, four, etc. hops away.
Therefore, our algorithm iterates on $\mR$ until the number of nonzeros remains the same (line 11 in Algorithm~\ref{alg1}).







\section{Communication Analysis}\label{sec:comm}

In this section we analyze the communication costs of diBELLA 2D and compare them with our 1D implementation.
First, we briefly consider the communication cost of the k-mer counting phase, which is common to both implementations.
Then we examine the communication costs of the overlapping phase and read exchange, which are the main differences between the two implementations.
The communication costs of the transitive reduction for the current implementation follow.
The communication costs are given in word count $W$ (\emph{bandwidth cost}) and number of messages $Y$ (\emph{latency cost}). 
The communication costs are summarized in Table~\ref{tab:comm} next to useful notations in Table~\ref{tab:params}.

\vspace{-.25em}
\subsection{Communication Cost of K-mer Counting}

The communication costs for this step depend on the properties of the input dataset and the settings of our algorithm, such as the depth $d$ of the dataset, the genome size $G$ (in nucleotides), the k-mer length $k$ and the read length $l$.


Our total input size is $G d \approx n l$.
Each processor has $(1/P)$th of the input. 
Each sequence has $(l-k+1)$ k-mers and each k-mer takes $k/4$ bytes using 2--bit compression per nucleotide.
Hence, the total size on each processor before communication is $n(l-k+1)k/4P$.
For long--read data $l-k+1 \approx l$, since $l$ is usually 2--3 orders of magnitude larger than $k$.  

The hash function maps k-mers uniformly and randomly on processors, so that each processor keeps $(1/P)$th of the data for itself and communicates the rest. 
For large $P$ we can assume $(P-1)/P \approx 1$ to avoid clutter. 
Hence, the bandwidth cost for k-mer counting per process is on average:

\begin{equation}
W = \frac{P-1}{P} \frac{n(l-k+1)k}{4P} \approx \frac{nlk}{4P} 
\end{equation}

Based on the available memory, we may have to perform several k-mer exchanges.
So the latency cost of k-mer counting is $Y = bP$, where $b$ is the batch count.

\vspace{-.25em}
\subsection{Communication Cost of Overlap Detection}

\begin{hide}
\Aydin{why is this first paragraph even relevant}
To study the communication costs of the overlapping phase, we use the parallel distributed memory communication model introduced by Ballard et al.~\cite{ballard2011minimizing}, which states that each processor has a local memory of size $M$ words such that we can store one copy of the output matrix across the processors.
This results in $M=\Omega(c^2 n/P)$ where $c$ is the sparsity of the output matrix, $n$ is the dimension of the output matrix, and $P$ is the number of processors.
\end{hide}


\begin{table}[t]
\caption{
Communication costs of diBELLA 1D and diBELLA 2D.
}
\centering
\begin{adjustbox}{width=\columnwidth}
{
  \begin{tabular}{|l|c|c|c|c|}
    \hline
    \multirow{2}{*}{\bf Task} &
      \multicolumn{2}{c|}{\bf Bandwidth} &
      \multicolumn{2}{c|}{\bf Latency} \\
      & {\bf diBELLA 1D} & {\bf diBELLA 2D} & {\bf diBELLA 1D} & {\bf diBELLA 2D}  \\
        \hline
        \hline
        K-mer Counting        & $nlk/4P$  &   $nlk/4P$  & ${bP}$  & ${bP}$ \\
        \rowcolor{Gray}
        Overlap Detection     & $a^2 m/P$ &  $a m/\sqrt{P}$   & $P$  & $\sqrt{P}$\\
        Read Exchange         & $cnl/P$ &   $2nl/\sqrt{P}$  & $\textrm{min}\{cnl/P, P\}$ & $\sqrt{P}$\\
        \rowcolor{Gray}
        Transitive Reduction  & - & $rn/\sqrt{P}$ & -  & $t \sqrt{P}$ \\
        \hline
  \end{tabular}
  }
  \end{adjustbox}
  \vspace{-1em}
\label{tab:comm}
\end{table}


 
 

In our application $\mA\mA\transpose$ is the output matrix $n \times n$, $\mA$ and $\mA\transpose$ are the input matrices of dimension $n \times m$ and $m \times n$ respectively.
The number of nonzeros in both $\mA$ and $\mA\transpose$ is $a m$, where $a$ is the density indicating the average number of sequences containing a particular k-mer.
The k-mer selection procedure that we present in~\cite{guidi2020bella} and that we also use in this work chooses an interval for the k-mer frequency, which in turn translates into the average number of sequences that can contain a given k-mer.

Our previous work, diBELLA 1D, computes overlap detection using distributed hash tables~\cite{ellis2019dibella}. 
K-mers are distributed to processors that allow them to detect candidate overlap pairs locally. 
This must be followed by a global reduction.
In terms of communication, this implementation is equivalent to a 1D sparse matrix multiplication using the outer product algorithm~\cite{buluc2008representation,ballard2013communication}.
The 1D outer product formulation distributes $\mA$ in block columns where the $i$th block column is denoted by $\mA_{:i}$, and $\mA\transpose$ in block rows where the $i$th block row is  denoted by $\mA\transpose_{i:}$. $\mC$ is distributed in block rows in diBELLA 1D. The computation can be written as
$ C = \sum_{i=1}^{P}{\mA_{:i} \mA\transpose_{i:}}$.

\begin{table}[t]
\caption{
List of symbols and annotations in our paper.
}
\centering
{
\centering
\begin{tabular}{|c|l|}
\hline
\textbf{Symbol} & \hspace{4em}\textbf{Description} \\ 
\hline
\hline
\emph{n} & Read set cardinality \\
\rowcolor{Gray}
\emph{m} & K-mer set cardinality \\
\emph{d} & Depth of coverage \\
\rowcolor{Gray}
\emph{k} & K-mer length \\
\emph{L} & Overlap length \\
\rowcolor{Gray}
\emph{l} & Read length \\
$\mA$ & Data matrix: reads--by--kmers \\
\rowcolor{Gray}
$\mC$ & Candidate overlap matrix: reads--by--reads \\
$\mR$ & Overlap matrix: reads--by--reads \\
\rowcolor{Gray}
$\mS$ & String matrix: reads--by--reads \\
\emph{a} & $\mA$ average density: $\sfrac{\rm nnz(\mA)}{m}$ \\
\rowcolor{Gray}
\emph{c}  & $\mC$ average density: $\sfrac{\rm nnz(\mC)}{n}$ \\
\emph{r} & $\mR$ average density: $\sfrac{\rm nnz(\mR)}{n}$ \\
\rowcolor{Gray}
\emph{s} & $\mS$ \hspace{.2em}average density: $\sfrac{\rm nnz(\hspace{.1em}\mS\hspace{.1em})}{n}$ \\
\emph{P} & Total number of processes \\
\rowcolor{Gray}
\emph{W} & Bandwidth cost \\
\emph{Y} & Latency cost \\
\hline
\end{tabular}
}
\vspace{-1em}
\label{tab:params}
\end{table}

Each k-mer exists on average in $a$ sequences, hence the local overlap detection $\mA_{:i} \mA\transpose_{i:}$ generates $a^2 m/P$ nonzeros on each processor.
These nonzeros must be reduced before performing pairwise alignment, so that no read-read pair is aligned more than a few (1-2) times, depending on the algorithm parameters.
This means that each processor exchanges $W_{1D} = a^2m/P$ words and the latency cost is $Y_{1D}=P$.

\begin{hide}
If a processor $P_j$ needs to calculate the $j$th block row, it needs to access the $j$th block column of $\mA$, $\mA_j$, already in its memory and the $j$th block row of $\mA\transpose$, $\mA\transpose_j$, on another processor.
A na\"ive block algorithm would communicate the entire block row of $\mA\transpose$, while the outer product implementation we use in the initial version of diBELLA determines exactly which rows of $\mA\transpose$ it needs to access (i.e. those containing some nonzeros) and communicates only those.
In practice, the outer product algorithm performs sparse communication.
In the outer product formulation, the $j$th column of $\mA$ and the $j$th row of $\mA\transpose$ are multiplied to form a rank-1 matrix.
The algorithm performs the same procedure for each row of $\mA\transpose$ and obtains $m$ different rank-1 matrices.
Finally it adds them together to obtain the resulting matrix $\mC$.

If we consider the block column $\mA_j$, we multiply it for the block row $\mA\transpose_j$, where both have width $m/P$.
Given the density \emph{a} of $\mA$, we can assume that there are $am/P$ rows of $\mA\transpose$ with at least one nonzero. 
If we consider that (a) the processor $P_j$ only needs to access columns of $\mA$ that correspond to nonzeros rows of $\mA\transpose$ and (b) each column of $\mA$ has $a$ nonzeros, we see that the number of nonzeros that the processor $P_j$ needs to access in $\mA$ is $W_{1D} = a^2 m/P$.
\end{hide}

In contrast, diBELLA 2D uses a 2D sparse matrix multiplication algorithm known as Sparse SUMMA~\cite{buluc2008challenges}.
The $P$ processors are logically organized in a $\sqrt{P}\times \sqrt{P}$ grid with row and column indexes, so that the $(i, j)$th processor is $P_{ij}$.
Each processor stores a $n/\sqrt{P}\times m/\sqrt{P}$ submatrix $\mA_{ij}$ and a $m/\sqrt{P}\times n/\sqrt{P}$ submatrix $\mA\transpose_{ij}$ in its local memory.
Each processor calculates a product of a block row of $\mA$ with a block column of $\mA\transpose$. 
Sparse SUMMA is an owner-computes algorithm, so we only need to consider the communication of input matrices.
If we assume a good load balance, which we achieve by randomly permuting k-mers and reads, $\mA_{ij}$ has $am/P$ nonzeros.
Each processor $P_{ij}$ receives $2 (\sqrt{P}-1) $ input blocks because $
\mC_{ij} = \sum_{i=k}^{\sqrt{P}}{\mA_{ik} \mA\transpose_{kj}}$.

For large $P$ we simplify $\sqrt{P}-1 \approx \sqrt{P}$ so that the number of nonzeros that a processor must collect is $W_{2D} = a m/\sqrt{P}$ and the latency cost is $Y_{2D} = \sqrt{P}$.

\vspace{-.25em}
\subsection{Communication Cost of Read Exchange}

The communication costs of the read exchange are derived from the analysis in the previous section.
The sequences are distributed to the processors by parallel I/O according to the corresponding implementation decomposition, i.e. 1D for our first implementation and 2D for the present work.

To compute the communication costs, we consider the candidate overlap matrix $\mC^{n\times n} = \mA\mA\transpose$. 
$\mC$ has $c n$ nonzeros (before computing pairwise alignment) where $c$ is its density per row or column, which indicates the average number of overlapping sequences for each read.
Each exchange costs $O(l)$.
The 1D algorithm exchanges at most $W_{1D} = c n l/P$ words and sends $Y_{1D} = \textrm{min}\{c n l/P, P\}$ messages, while the current 2D implementation exchanges at most $W_{2D} = 2 n l/\sqrt{P}$ words and sends $Y_{2D} = \sqrt{P}$ messages.

diBELLA 1D communicates at most one read per nonzero because an alignment task is only assigned to a processor if this processor has at least one of the two sequences involved.
diBELLA 2D communicates the full range of sequences that a processor may need and starts the read exchange immediately after the initial data partition, so that communication overlaps with computation.

The 1D algorithm has better scaling with increased concurrency, but has a large constant $c$, whose typical value often exceeds $1000$ for large genomes, as shown in Table~\ref{tab:sparsities}. 
To overcome this large constant and communicate fewer words than the 2D algorithm, the 1D algorithm would require $(c^2/4)$--way parallelism. 
Ellis et al.~\cite{ellis2019dibella} show that $c \approx 2d$ for a perfect overlapper.
In practice, $c$ is much larger than $d$ and $c/2d$ can be considered the \emph{inefficiency factor} of an overlapper.

\begin{hide}
In  we report the experimental values for diBELLA 2D for the input we use for the evaluation. 
If we consider $P \to \infty$ and a given input, diBELLA 1D would scale better than diBELLA 2D because it would perform sparser communication~\cite{ballard2013communication}.
The 2D algorithm can achieve near linear scaling until bandwidth costs begin to dominate and then scaling proportionally to $\sqrt{P}$.
\end{hide}

\begin{table}[t]
\caption{
List of experimental values of sparsity for diBELLA 2D.
}
\centering
\begin{adjustbox}{width=\columnwidth}
{
\centering
\begin{tabular}{|l|c|c|c|c|}
\hline
\textbf{Dataset} & \textbf{Depth (\textit{d})} & \textbf{$\mC$ density (\textit{c})} & \textbf{Inefficiency ($\textbf{\sfrac{c}{2d}}$)} & \textbf{$\mR$ density (\textit{r})}  \\ 
\hline
\hline
E. coli & 30 &\hspace{1em}145.9 & \hspace{.5em}2.4 & 6.4 \\
\rowcolor{Gray}
C. elegans &  40 & 1,579.7 & 19.7 & 8.1 \\
H. sapiens & 10 & 1,207.7 & 60.4 & 1.3 \\
\hline
\end{tabular}
}
\end{adjustbox}
\vspace{-.5em}
\label{tab:sparsities}
\end{table}
\begin{table}[t]
\caption{
Data sets used during evaluation: name, depth, number of sequences in the input, average read length, input size, genome size, and error rate.
}
\centering
\begin{adjustbox}{width=\columnwidth}
{
\centering
\begin{tabular}{|l|c|r|r|c|r|c|}
\hline
\textbf{Label} &  \textbf{Depth} & \textbf{Reads (K)} & \textbf{Length} & \textbf{Input (GB)} & \textbf{Size (Mb)}  & \textbf{Error} \\ 
\hline
\hline
C. elegans & 40 & 420.7 & 11,241 & \hspace{0.5em}4.8 & 100 & 0.13 \\
\rowcolor{Gray}
H. sapiens & 10 & 4,421.6 & 7,401 & 33.1 & 3,000 & 0.15 \\
\hline
\end{tabular}
}
\end{adjustbox}
\vspace{-1em}
\label{tab:data}
\end{table}

\vspace{-.25em}
\subsection{Communication Cost of Transitive Reduction}

The sparse matrix multiplication dominates the runtime of the transitive reduction algorithm.
The communication costs for the squaring of $\mR$ follow from the previous analysis and are $W_{2D} = r n/\sqrt{P}$, where $r\leq c$ is the sparsity of the overlap matrix $\mR^{n\times n}$ after performing the pairwise alignment, which often leads to the discarding of nonzeros, and $Y_{2D} = \sqrt{P}$.
The transitive reduction algorithm also contains some element-wise sparse routines, but these are executed in-place so that they do not contribute to communication time.
While the transitive reduction loop is repeated until convergence, the number of iterations is often a small constant (denoted $t$ in Table~\ref{tab:comm}) and the geometrically decreasing density after each iteration makes the total communication volume asymptotically equal to that of the first iteration.

\section{Experimental Setup}\label{sec:setup}

\begin{table*}[t]
\caption{Details of the machines used for evaluation: name, number of physical cores per node, processor max turbo frequency, processor model, memory, network, and L1, L2, L3 caches sizes.
}
\begin{adjustbox}{width=\textwidth}
{
\centering
\begin{tabular}{|l|c|c|l|c|c|r|r|r|}
\hline
\textbf{Platform} & \textbf{Cores/Node} & \textbf{Frequency (GHz)} &
\textbf{Processor} & \textbf{Memory (GB)} & \textbf{Network and Topology} & \textbf{L1}\hspace{.65em} & \textbf{L2}\hspace{.65em}  & \textbf{L3}\hspace{.65em}  \\ 
\hline
\hline
Cori Haswell & 32 & 
3.6 & Intel Xeon E5-2698V3 & 128 & Aries Dragonfly & 64KB & 256KB & 40MB \\
\rowcolor{Gray}
Summit CPU & 42 &  4.0 & IBM POWER9  & 512 & InfiniBand Non-Blocking Fat Tree  & 32KB & 512KB & 10MB\\
\hline
\end{tabular}
}
\end{adjustbox}
\label{tab:machines}
\end{table*}

Our experiments were performed on two machines: the Cray XC40 supercomputer Cori at NERSC and the IBM  supercomputer Summit at Oak Ridge National Laboratory.
On Cori we use the Haswell partition, while on Summit we use only IBM POWER9 CPUs.
Using two architectures shows that our algorithm scales on different architectures.
However, this is not intended to be a cross--platform comparison, as our algorithm is not specifically optimized for either platform.

Each Haswell node on the Cori system consists of two 2.3 GHz 16-core Intel Xeon processors and has a total memory of 128 GB.
Each Summit node has two 22-core IBM POWER9 processors and 512 GB DDR4 of RAM.
Because one core per Summit half-node is reserved for OS, each node has a maximum of 42 cores available for application codes. 
In this paper, we do not utilize the GPUs available on Summit. 
For details on the two architectures, see Table~\ref{tab:machines}.

To investigate the parallel performance of our algorithm, we use two data sets from Pacific Biosciences (CLR technology)~\cite{rhoads2015pacbio} with different sizes:
Caenorhabditis elegans (C. elegans) and Homo sapiens (H. sapiens).
Details of the two data sets are given in Table~\ref{tab:data}.
Our algorithm is also suitable for other long--read technologies such as Pacific Biosciences CCS~\cite{wenger2019accurate} (or HiFi) and Oxford Nanopore~\cite{jain2016oxford}.
In this paper we only run with CLR data as our parameters are tuned to it and the accuracy of our tool for CLR input is reported in the single node BELLA paper~\cite{guidi2020bella}.

The experiments are divided into two groups: (a) parallel performance and scalability of diBELLA 2D and (b) performance comparison with related work.
In the latter case, we compare the overlap detection of diBELLA 2D with diBELLA 1D~\cite{ellis2019dibella} and minimap2~\cite{li2018minimap2} written in C for shared memory, while we compare our transitive reduction algorithm with SORA~\cite{paul2018sora} written in Scala on top of Apache Spark~\cite{zaharia2016apache}.

diBELLA 2D and 1D run with the same input and alignment setting, i.e. $k = 17$ and maximum k-mer frequency equal to 4, while minimap2 run with its default setting for CLR data.
The results from minimap2 and SORA were only collected from Cori, as minimap2 uses SSE intrinsics that are not supported on the IBM POWER9 processor, and Summit has no support for Apache Spark.
For minimap2 we only report single node performance because it does not implement distributed memory parallelism.
To compare transitive reduction, we used the output of diBELLA 2D as input for SORA, which is an overlap graph consisting of 5.8M edges and 4.4M vertices for the H. sapiens data set and 4.2M edges and 0.4M vertices for C. elegans.
We only compare the execution time of the transitive reduction by removing all start and shutdown times from Apache Spark and the time dedicated to I/O.

On Cori, we used \texttt{gcc-8.3.0} and the \texttt{O3} flag to compile C/C++ codes, while on Summit we used \texttt{gcc-8.1.1}.
On both Cori and Summit, we used the default MPI implementation.
SORA used \texttt{jdk/1.8.0\_202} and \texttt{spark/2.3.0}.
In the next section we report the average runtime over 10 runs for each experiment, except for the H. sapiens data set at low concurrency, where we report the average over three runs. 


\section{Experimental Results}\label{sec:results}


\begin{figure}[t]
    \centering
    \includegraphics[width=\columnwidth]{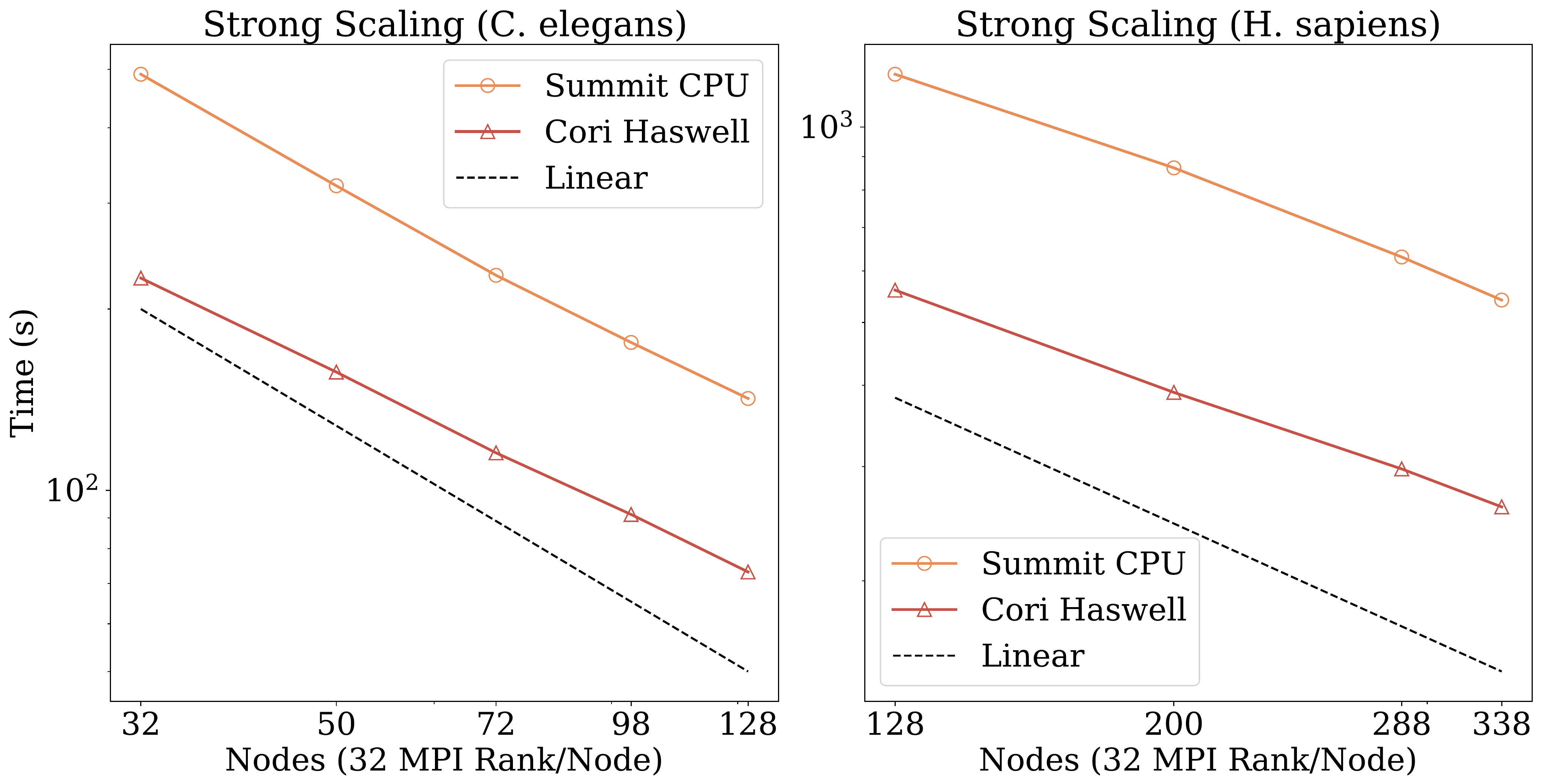}
    \caption{diBELLA 2D strong scaling on Cori Haswell and Summit CPU using 32 MPI rank/node on C. elegans (left) and on H. sapiens (right).}
    \vspace{-1em}
    \label{fig:strong_scaling}
\end{figure}


In this section we first examine the parallel performance of diBELLA 2D and its individual components and then compare our performance with the state of the art.

\subsection{Detailed Analysis of diBELLA 2D}

Figure~\ref{fig:strong_scaling} illustrates the strong scaling of our algorithm for C. elegans on the left and for H. sapiens on the right.
In this figure, the two machines run on $P=$ \{32, 72, 128\} nodes using 32 MPI ranks/node for C. elegans and $P=$ \{128, 200, 288, 338\} for H. sapiens. 
For C. elegans, diBELLA 2D achieves a parallel efficiency of 83\% on Summit CPU, while it achieves a parallel efficiency of 68\% on Cori Haswell.
For H. sapiens, the parallel efficiency of both machines is over 80\% with a peak of 92\% on Summit CPU.
These results show the near linear scaling behavior of our overall algorithm using a large input on two different architectures.

\begin{figure}[t]
    \centering
    \includegraphics[width=\columnwidth]{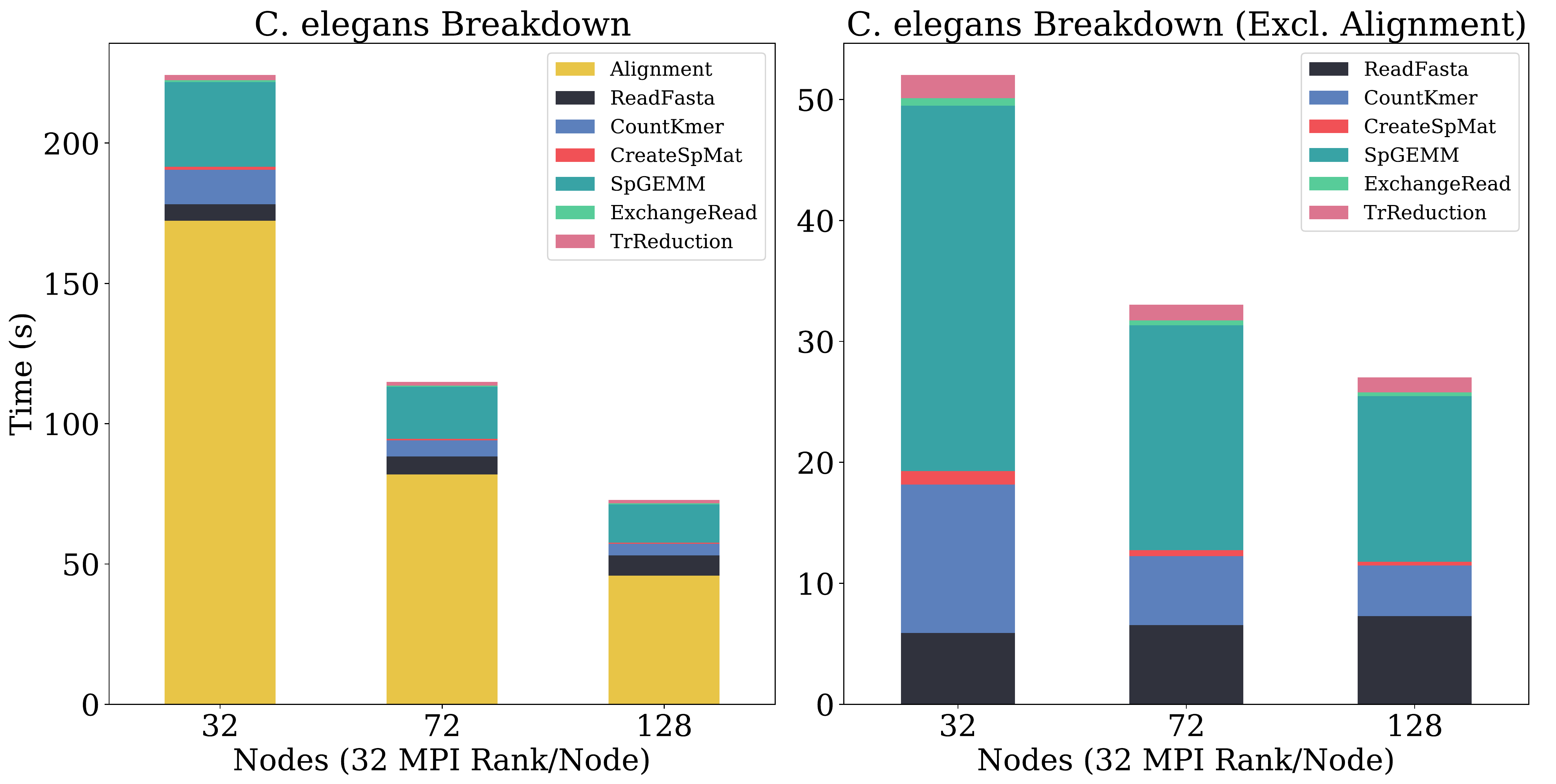}
    \caption{diBELLA 2D runtime breakdown on Cori Haswell (C. elegans).}
    \vspace{-.5em}
    \label{fig:cori_haswell_celegans}
\end{figure}

\begin{figure}[t]
    \centering
    \includegraphics[width=\columnwidth]{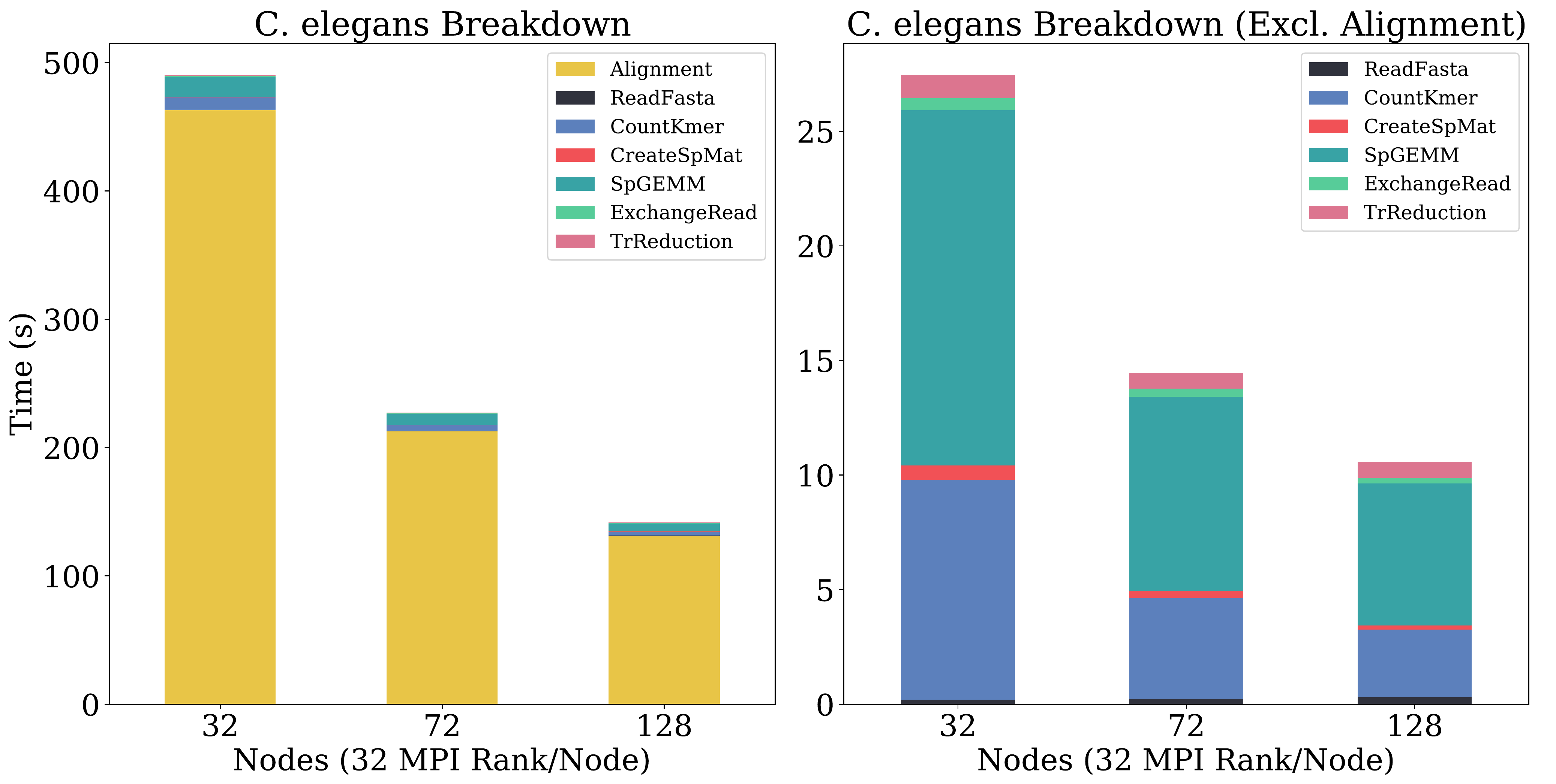}
    \caption{diBELLA 2D runtime breakdown on Summit CPU (C. elegans).}
    \vspace{-1em}
    \label{fig:summit_celegans}
\end{figure}

\begin{figure}[t]
    \centering
    \includegraphics[width=\columnwidth]{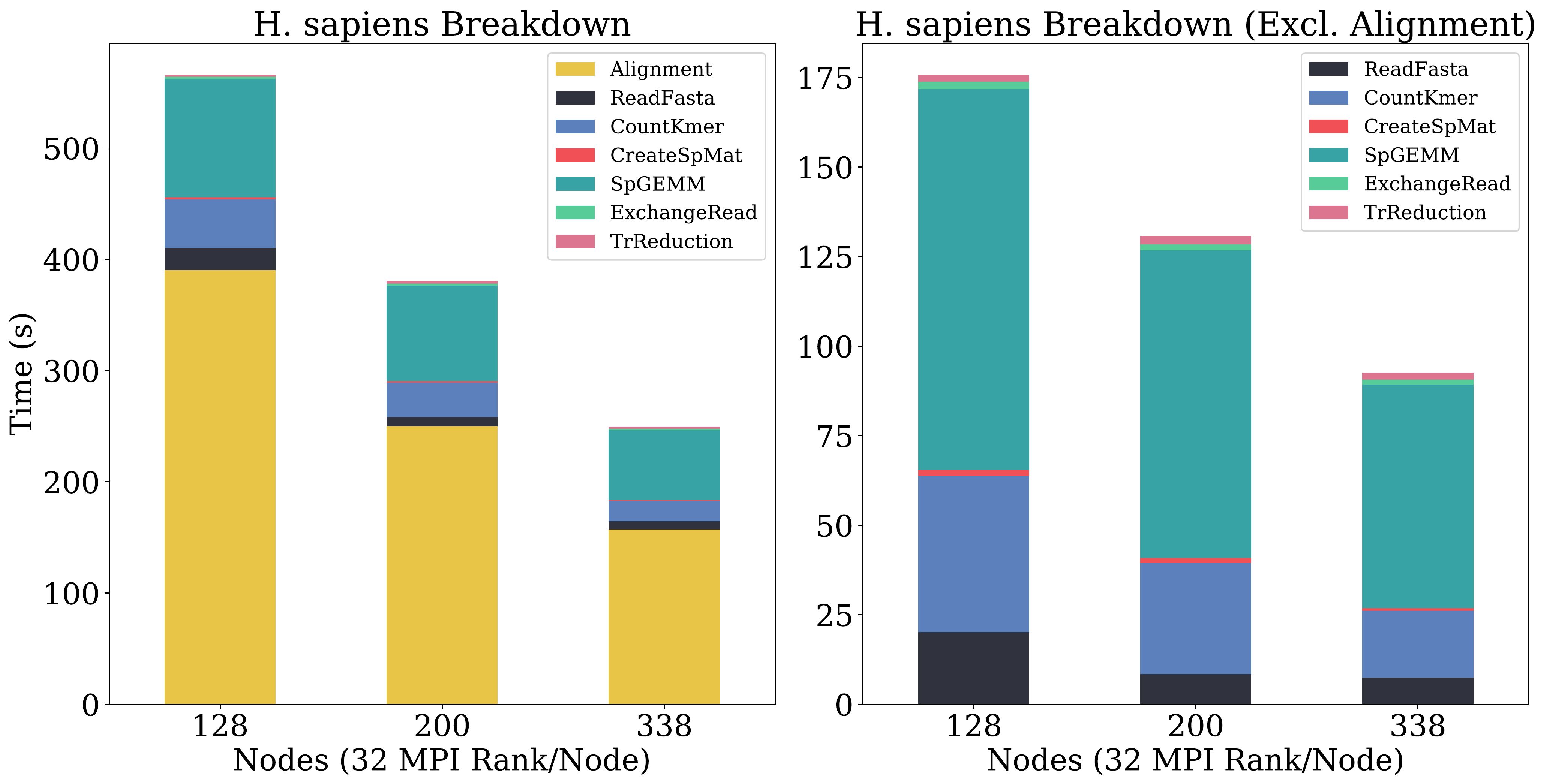}
    \caption{diBELLA 2D runtime breakdown on Cori Haswell (H. sapiens).}
    \vspace{-.5em}
    \label{fig:cori_haswell_human}
\end{figure}

\begin{figure}[t]
    \centering
    \includegraphics[width=\columnwidth]{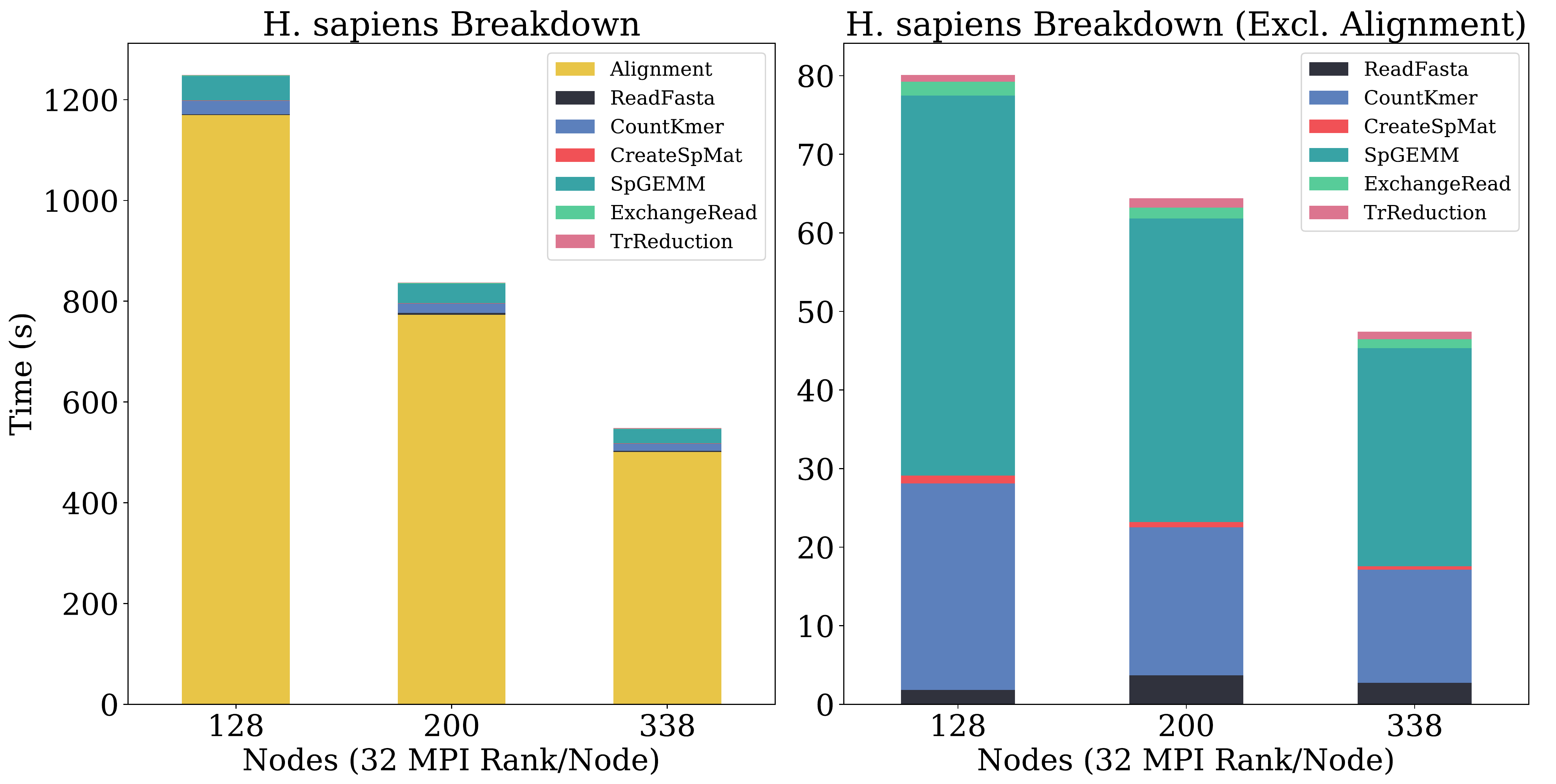}
    \caption{diBELLA 2D runtime breakdown on Summit CPU (H. sapiens).}
    \vspace{-1em}
    \label{fig:summit_human}
\end{figure}

Figures~\ref{fig:cori_haswell_celegans}--\ref{fig:summit_celegans} (C. elegans) and Figures~\ref{fig:cori_haswell_human}--\ref{fig:summit_human} (H. sapiens) show the runtime breakdown of diBELLA 2D on the two machines.
In each breakdown, the plot on the left shows total execution time including pairwise alignment, while the plot on the right excludes pairwise alignment.
We included plots excluding pairwise alignment because alignment takes a large proportion of the runtime and makes it difficult to see the scaling of the other stages.
From bottom to top the layers are ordered according to the legend.
The first layer is the pairwise alignment, i.e. the time needed to align all non-zero pairs in the candidate overlap matrix $\mC$.
\texttt{ReadFastq} is the time spent reading and parsing the input file in parallel.
Immediately after this step, we start exchanging sequences to overlap this communication with the subsequent computation.
\texttt{CountKmer} corresponds to the k-mer counting stage, and \texttt{CreateSpMat} corresponds to the time needed to create the input matrices $\mA$ and $\mA\transpose$.
\texttt{SpGEMM} includes both the communication time and the computation time to create the candidate overlap matrix $\mC = \mA\mA\transpose$.
\texttt{ExchangeRead} times the period from the end of \texttt{SpGEMM} until all sequence exchanges are complete. 
Depending on the MPI implementation, the read exchange may potentially overlap with k-mer counting and overlap detection phases. 
Finally, \texttt{TrReduction} is the transitive reduction time.

Figures~\ref{fig:cori_haswell_celegans}--\ref{fig:summit_celegans} show diBELLA 2D performance breakdown on the C. elegans dataset using $P{=}\{32,72,128\}$ nodes of each machine.
diBELLA 2D runs faster overall on Cori Haswell than on Summit CPU.
The relative proportion of pairwise alignment in the total runtime increases on Summit CPU compared to Cori Haswell.
SeqAn's pairwise alignment is probably not optimized for IBM processors, so we refrain from making architecture comparisons based solely on this data.

On the two machines, the sparse matrix computation $\mA\mA\transpose$ for overlap detection after pairwise alignment has the greatest contribution to runtime.
The parallel read I/O does not show any scaling and its performance deteriorates when the number of processes is increased.
For C. elegans, overlap detection has a parallel efficiency of 55\% for Cori Haswell and 63\% for Summit CPU, while k-mer counting (consisting of first and second passes) has a parallel efficiency of 70--80\% across machines.
The read exchange has a parallel efficiency of 50\% across machines. 
Despite a parallel efficiency of 38\% on Cori Haswell and Summit CPU, our transitive reduction shows a significant speedup compared to a competing distributed memory implementation. 

Figures~\ref{fig:cori_haswell_human}--\ref{fig:summit_human} tell a similar story for H. sapiens.
The size of the data set, which is about 10$\pmb\times$ of C. elegans, mitigates the contribution of I/O to the overall runtime.
The parallel efficiency of $\mA\mA\transpose$ increases to 65\% on both machines.
The parallel efficiency of k-mer counting reaches 89\% on Cori Haswell.
The formation of $\mA$ and $\mA\transpose$ has negligible impact on the total runtime, yet it scales almost linearly with a parallel efficiency of over 80\% on both machines and data sets.
\begin{hide}
The parallel efficiency of our algorithm improves with increasing problem size.
\end{hide}

\subsection{Comparison with the State of the Art}

To demonstrate the competitiveness of our approach, we now compare it with prior work in the literature.

\begin{figure}[t]
    \centering
    \includegraphics[width=\columnwidth]{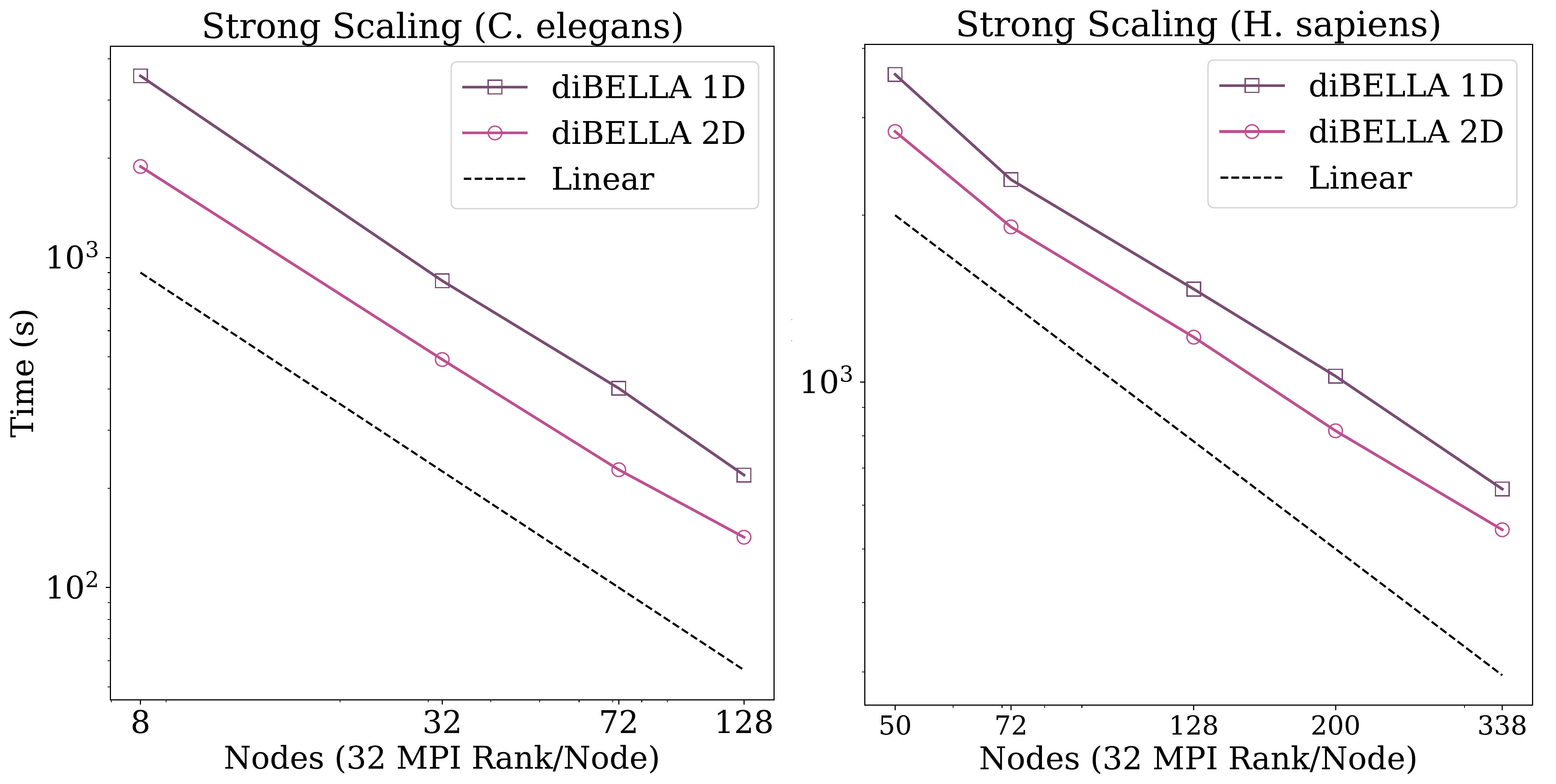}
    \caption{Comparison of diBELLA 2D and diBELLA 1D~\cite{ellis2019dibella} on Summit CPU.
    }
    \vspace{-1em}
    \label{fig:dibella1d}
\end{figure}

First, we compare the overall runtime and scaling of diBELLA 2D with diBELLA 1D~\cite{ellis2019dibella}, subtracting the transitive reduction time from diBELLA 2D, since the 1D version does not implement this step.
This comparison was made on Summit CPU and is shown in Figure~\ref{fig:dibella1d} for C. elegans and H. sapiens.
diBELLA 2D and 1D differ mainly in the way they perform overlap detection and communicate sequences before pairwise alignment.
They exhibit similar near linear scaling behavior, but diBELLA 2D consistently outperforms the 1D implementation by $1.5\mbox{--}1.9\pmb\times$ (average $1.7\pmb\times$) for the C. elegans data set and $1.2\mbox{--}1.3\pmb\times$ (average $1.2\pmb\times$) for H. sapiens.

For completeness, we evaluate diBELLA 2D against minimap2~\cite{li2018minimap2}, a popular shared memory algorithm for overlap detection.
To do this, we run minimap2 on a single node using 32 OpenMP threads and compare its runtime to diBELLA 2D on a different number of nodes using 32 MPI ranks/node. 
It is important to note that minimap2 and diBELLA 2D perform significantly different computations.
In particular, minimap2 does not perform base-level pairwise alignment and instead estimates pairwise similarity from the number of shared minimizers, making it significantly faster.
Nevertheless, they are ultimately aimed at solving the same problem, which is why we provide a comparison here.
For the runtimes of diBELLA 2D we refer to Figure~\ref{fig:strong_scaling}.
For C. elegans, minimap2 is about 2$\pmb\times$ faster than diBELLA 2D at $P =$ 8, while at higher concurrency diBELLA 2D is 1.6$\pmb\times$, 3.2$\pmb\times$, and 5$\pmb\times$ faster than minimap2.
The speedup of diBELLA 2D over minimap2 is 9.5$\pmb\times$, 13.7$\pmb\times$, and 20.6$\pmb\times$ at $P{=}\{128, 200, 338\}$ for the H. sapiens dataset.

Finally, we compare our transitive reduction algorithm with the transitive reduction module of SORA~\cite{paul2018sora}, a distributed memory implementation of transitive reduction  from overlap graph to string graph based on Apache Spark and GraphX.
Our transitive reduction algorithm is currently integrated into our pipeline, so for fairness reasons we do not include startup, shutdown, and I/O time for SORA.
The input of SORA is the overlap matrix $\mR$ of diBELLA 2D, therefore the two transitive reduction algorithms work with the same input.
The results are summarized in Table~\ref{tab:apache}.
Our transitive reduction algorithm has a speedup of up to 29$\pmb\times$ for C. elegans and up to 13.3$\pmb\times$ for H. sapiens.

\newcolumntype{g}{>{\columncolor{Gray}}c}
\begin{table}[t]
\caption{
Comparison of transitive reduction (in seconds) between diBELLA 2D and SORA~\cite{paul2018sora} on Cori Haswell.
}
\centering
\begin{adjustbox}{width=.8\columnwidth}
{
\centering
\begin{tabular}{|l|g|g|g|g|}
\hline
\rowcolor{white}
\textbf{Dataset} & \textbf{Nodes} & \textbf{SORA} & \textbf{diBELLA 2D} & \textbf{Speed-Up} \\ 
\hline
\hline
\rowcolor{white}
\multirow{3}{*}{C. elegans} & 32 & 34.6 & 1.9 &  18.2$\times$ \\
& 72 & 34.3 & 1.3 &26.4$\times$ \\
\rowcolor{white}
& 128 & 34.9 & 1.2 & 29.0$\times$ \\
\hline
\hline
\rowcolor{white}
\multirow{3}{*}{H. sapiens} & 128 & 23.4 & 1.9 &  12.4 $\times$ \\
& 200 & 24.3 & 2.3 &  10.5 $\times$ \\
\rowcolor{white}
& 338 & 25.3 & 1.9 & 13.3  $\times$ \\
\hline
\end{tabular}
}
\end{adjustbox}
\vspace{-1em}
\label{tab:apache}
\end{table}
\section{Conclusions}\label{sec:conclusions}

In this work, we presented a scalable distributed memory approach called diBELLA 2D for overlap detection and layout simplification in the context of \emph{de novo} genome assembly. 
diBELLA 2D relies on linear algebraic operations over semirings using 2D distributed sparse matrices. Using sparse matrices for both overlap detection and transitive reduction reduces the need for different data structures normally used in genome assembly. Well-studied distributed-memory algorithms on sparse matrices, such as sparse matrix-matrix multiplication, allows diBELLA to efficiently parallelize the computation without losing expressiveness, thanks to the semiring abstraction.
We provided detailed communication analysis for parallel overlap detection and transitive reduction, which were missing from previous work. In particular, our analysis shows that the new 2D overlap detection algorithm reduces communication compared to the existing 1D algorithm for commonly utilized concurrencies in the range of $100\mbox{--}10000$ processors. This translates into a speedup of $1.2\mbox{--}1.9\pmb\times$ in our experiments. For transitive reduction, our approach is $10\mbox{--}29\pmb\times$ faster than an existing distributed-memory implementation.

Our approach paves the way for high performance assembly of large genomes where it is impractical to assemble them in shared memory or on a small cluster.
Fast and efficient \emph{de novo} assembly enables the study of previously uncharacterized genomes~\cite{simpson2012efficient}.
Deciphering the nucleotide sequence is also of crucial importance when a reference genome is available. 
In mapping sequences to a reference, individual--specific genetic variations are often lost. 
But they are crucial, for example, to discover disease--causing mutations~\cite{chaisson2015genetic}.

Future work includes reducing memory consumption so that diBELLA 2D can assemble large genomes at low concurrency if desired. 
For this purpose, we can form only a part of the candidate overlap matrix in each time step, aligning only sequences belonging to this part, and removing the spurious entries before moving on to the next region of the output matrix.
Also, we plan to use GPUs in both pairwise alignment and k-mer counting to boost the performance of our pipeline.

\section*{Acknowledgments}

This work is supported by the Advanced Scientific Computing Research (ASCR) program within the Office of Science of the DOE under contract number DE-AC02-05CH11231. We used resources of the NERSC supported by the Office of Science of the DOE under Contract No. DEAC02-05CH11231. This research was also supported by the Exascale Computing Project (17-SC-20-SC), a collaborative effort of the U.S. Department of Energy Office of Science and the National Nuclear Security Administration.
This research also used resources of the Oak Ridge Leadership Computing Facility at the Oak Ridge National Laboratory, which is supported by the Office of Science of the U.S. Department of Energy under Contract No. DE-AC05-00OR22725.
Thanks to Alok Tripathy for useful suggestions and valuable discussions.


\bibliographystyle{IEEEtran}
\bibliography{refs}

\end{document}